\documentclass[reprint,prl,twoside,twocolumn,floatfix,superscriptaddress,noshowpacs]{revtex4-1}
\usepackage{times}
\usepackage{graphicx}
\usepackage{dcolumn}
\usepackage{bm}
\usepackage{epsfig}
\usepackage{amsmath}
\usepackage{cases}
\usepackage{upgreek}
\usepackage{amssymb}
\usepackage{wasysym}
\usepackage{float}

\begin{document}

\title{Microscopic Parameters from High-Resolution Specific Heat Measurements \\ on Overdoped BaFe$\bm{_{2}}$(As$\bm{_{1-x}}$P$\bm{_{x}}$)$\bm{_{2}}$ Single Crystals}

\author{Z. Diao}
\author{D. Campanini}
\affiliation{Department of Physics, Stockholm University, AlbaNova University Center, SE\,-\,106 91 Stockholm, Sweden}
\author{L. Fang}
\affiliation{Department of Chemistry, Northwestern University, IL 60208, USA}
\affiliation{Materials Science Division, Argonne National Laboratory, 9700 South Cass Avenue, IL 60439, USA}
\author{W.-K. Kwok}
\affiliation{Materials Science Division, Argonne National Laboratory, 9700 South Cass Avenue, IL 60439, USA}
\author{U. Welp}
\affiliation{Materials Science Division, Argonne National Laboratory, 9700 South Cass Avenue, IL 60439, USA}
\author{A. Rydh}\thanks{Author to whom correspondence should be addressed. Electronic address: andreas.rydh@fysik.su.se}
\affiliation{Department of Physics, Stockholm University, AlbaNova University Center, SE\,-\,106 91 Stockholm, Sweden}

\date{\today}

\begin{abstract}
We investigate the electronic specific heat of overdoped BaFe$_{2}$(As$_{1-x}$P$_{x}$)$_{2}$ single crystals in the superconducting state using high-resolution nanocalorimetry. From the measurements, we extract the doping dependence of the  condensation energy, superconducting gap $\Delta$, and related microscopic parameters. We find that  the anomalous scaling of the specific heat jump $\Delta C \propto T_{\mathrm{c}}^3$, found in many iron-based superconductors, in this system originates from a $T_\mathrm{c}$-dependent ratio $\Delta/k_\mathrm{B}T_\mathrm{c}$ in combination with a doping-dependent density of states $N(\varepsilon_\mathrm{F})$. A clear enhancement is seen in the effective mass $m^{*}$ as the composition approaches the value that has been associated with a quantum critical point at optimum doping. However, a simultaneous increase in the superconducting carrier concentration $n_\mathrm{s}$ maintains the superfluid density, yielding an apparent penetration depth $\lambda$ that decreases with increasing $T_\mathrm{c}$ without sharp divergence at the quantum critical point. Uemura scaling indicates that $T_\mathrm{c}$ is governed by the Fermi temperature $T_\mathrm{F}$ for this multi-band system.

\end{abstract}


\maketitle


Iron-based superconductors provide new grounds for investigating unconventional pairing mechanisms in high-$T_\mathrm{c}$ superconductivity, a long-standing mystery in modern condensed matter physics. Among the large number of compounds discovered in this family, the isovalently doped BaFe$_{2}$(As$_{1-x}$P$_{x}$)$_{2}$ (P-122) system \cite{Jiang:2009ft} has attracted widespread attention. An advantage with the P-122 system is that crystals on the strongly overdoped side are clean enough to display the de Haas-van Alphen effect (dHvA), yielding information about the band structure that can be directly compared with calculations \cite{Analytis:2010hl,Shishido:2010fn,Carrington:2011gl,Arnold:2011ip}. In the system, several methods indicate the presence of a possible quantum critical point (QCP) near optimum doping $x_\mathrm{cr}\simeq 0.3$ \cite{Hashimoto:2012, Walmsley:2013, Analytis:2014, Shibauchi:2014,Putzke:2014}. In particular, the effective mass $m^\star$ of at least one of the electron pockets at the X-point is enhanced when approaching $x_\mathrm{cr}$  \cite{Shishido:2010fn, Walmsley:2013}. Recently, it was reported that the London penetration depth $\lambda_\mathrm{L}$ also has a maximum at $x_\mathrm{cr}$ \cite{Hashimoto:2012}. This is unexpected, since this corresponds to a minimum in the superfluid density $\rho _{s}$ when $T_\mathrm{c}$ has its maximum, in contrast to the Uemura scaling $T_\mathrm{c} \propto n_\mathrm{s}/m^\star \propto \rho _{s}=\lambda _\mathrm{L}^{-2}$ of cuprate and other high-$T_\mathrm{c}$ superconductors \cite{Uemura:1991, Uemura:2004}. One of the important pathways to address this discrepancy as well as to gain further understanding of the pairing mechanism is through specific heat measurements as thermodynamic properties are closely linked to the nature of the superconducting state. It is known that the thermodynamics of the superconducting state of P-122 is anomalous in certain ways. The absolute specific heat jump at $T_\mathrm{c}$ is found to follow the so-far unexplained BNC scaling, $\Delta C(T_c) \propto T_c^3$, seen for many iron-based superconductors \cite{Budko:2009ff, Chaparro:2012dw}. However, thermodynamic evidence of the anomalous behavior in $\lambda_\mathrm{L}$ around $x_\mathrm{cr}$ has so far been rather ambiguous \cite{Walmsley:2013} and it is still debated which anomalies are central to the pairing mechanism.

Here we perform high-resolution specific heat measurements of overdoped BaFe$_{2}$(As$_{1-x}$P$_{x}$)$_{2}$ single crystals using a nanocalorimetry system capable of providing good absolute accuracy on very small samples. From the specific heat data we extract the doping dependence of the condensation energy $\Delta F(0)$, superconducting gap $\Delta (0)$, density of states at the Fermi level $N(\varepsilon _\mathrm{F})$ as well as microscopic parameters, i.e., coherence length $\xi _\mathrm{ab}(0)$ and penetration depth $\lambda _\mathrm{ab}(0)$ (including the related parameters $\xi_0$ and $\lambda_\mathrm{L}$). We find that the anomalous BNC scaling may be explained by a joint effect of the doping dependences of $\Delta (0)/k_{\mathrm{B}}T_{\mathrm{c}}$ and $N(\varepsilon _{\mathrm{F}})$. While a significant $m^{\star}$ enhancement is observed close to optimum doping, it does not lead to a drop in the superfluid density due to a simultaneous increase in the superconducting carrier concentration. The evolution of $T_{\mathrm{c}}$ scales with the Fermi temperature $T_{\mathrm{F}}$, in accordance with the Uemura relation for cuprates and several heavy-fermion systems. This strongly suggests that overdoped BaFe$_{2}$(As$_{1-x}$P$_{x}$)$_{2}$ belongs to this group of unconventional superconductors with an electronic origin of the coupling mechanism, governed by $T_\mathrm{F}$.

High-purity BaFe$_{2}$(As$_{1-x}$P$_{x}$)$_{2}$ crystals with $x=0.32$, 0.50 and 0.55 were grown using a self-flux method \cite{Fang:2011,Chaparro:2012dw}. Small crystals with ${\sim}150\,\upmu\mathrm{m}$ side were selected and gently cleaved to obtain plate-like samples with shiny surfaces. Specific heat was measured using a differential, auto-adjusting nanocalorimeter operating in constant-phase ac mode to yield combined high resolution and good absolute accuracy \cite{Tagliati:2012gm,Tagliati:2011be}. The sample temperature oscillation amplitude was set to 0.5\% of the absolute temperature. Device and thermal adhesive grease addenda were measured separately. To determine the absolute specific heat, the crystal volumes were estimated from scanning electron and optical microscopy. The resulting values at maximum $C/T$, where the phonon contribution dominates, were found within 10\% agreement with literature values on the parent compound BaFe$_{2}$As$_{2}$ \cite{Rotter:2008} as well as non-superconducting Ba(Fe$_{0.88}$Mn$_{0.12}$)$_{2}$As$_{2}$ crystals \cite{Popovich:2010}.


\begin{figure*}[t]
\begin{center}
\includegraphics[width=0.98\linewidth]{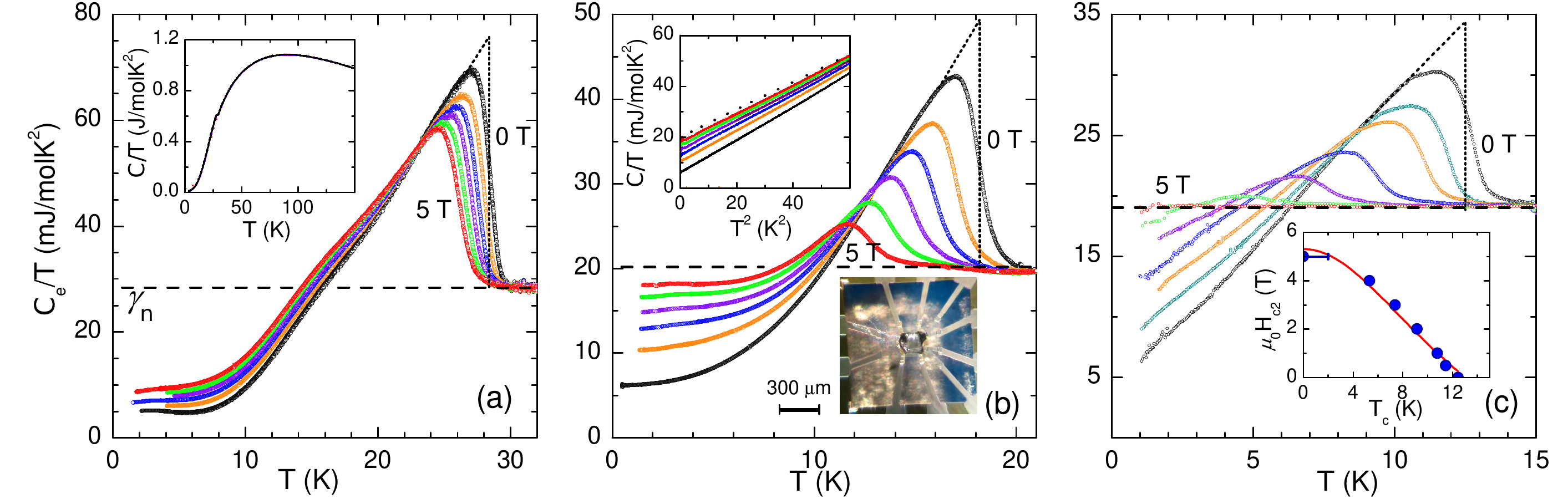}
\end{center}
\caption{(Color online) Temperature dependence of the electronic specific heat as $C_\mathrm{e}/T$ for BaFe$_{2}$(As$_{1-x}$P$_{x}$)$_{2}$ samples in fields from 0 to 5~T for (a) $x = 0.32$, (b) $x = 0.50$, and (c) $x = 0.55$. Field steps are 1 T except in (c) where 0.5~T is also shown. Dotted lines illustrate the entropy-conserving construction to determine $T_\mathrm{c}$. Inset of (a) shows the total specific heat for the $x=0.32$ sample (shown as $C/T$). Upper inset of (b) shows the low-temperature specific heat as $C/T$ vs $T^2$ for the same conditions. The dashed line represents the normal state. Lower inset of (b) shows a sample mounted on the calorimeter. Inset of (c) shows the upper critical field $\mu_{0}H_\mathrm{c2}$ for the $x=0.55$ sample.}
\label{Fig1}
\end{figure*}

Figure~\ref{Fig1} shows the electronic specific heat for three samples of different composition in magnetic fields $\mathbf{H} \parallel \mathbf{c}$. Comparing the zero-field temperature dependences of panel (a) through (c), there seems to be a tendency of going from a fully gapped system for $x=0.32$ to a system with pronounced gap anisotropy and possible nodes for $x=0.55$. However, the magnetic field dependence also indicates a faster-than-linear increase of $\gamma(H)$ for all samples. This supports the conclusion of thermal conductivity and penetration depth measurements that possible accidental nodes reside on the high Fermi velocity electron sheets \cite{Yamashita:2011} to which specific heat measurements are not as sensitive. Even for $x=0.32$ there is also a significant residual specific heat coefficient $\gamma_\mathrm{r} \sim 5\,\mathrm{mJ/molK}^{2}$, constituting roughly $17\%$ of the normal state Sommerfeld coefficient $\gamma_n$. Here, $\gamma_\mathrm{r}$ is largely invariant with doping, in contrast to Co-122 where an increase in $\gamma_\mathrm{r}$ away from optimal doping can be clearly identified \cite{Hardy:2010hr}. 

The phonon background of the $x=0.32$ and $x=0.50$ crystals were obtained by applying a simple Debye law and entropy conservation on the partially field-suppressed states. The normal state of the $x=0.50$ sample is illustrated as a dashed line in the inset of Fig.~\ref{Fig1}(b). A $5\,\mathrm{T}$ magnetic field is sufficient to fully suppress the superconducting state for the $x=0.55$ sample, as indicated in the bottom inset of Fig.~\ref{Fig1}(c). The Debye temperatures of all three samples are within 10\% of each other.

\begin{table}[t]
   \caption{\label{Table1} Parameters from specific heat using the specified relations and assumptions.}  
  \begin{ruledtabular}
  \begin{tabular}{lcccc}
   Parameter & unit & $x=0.32$ & $0.50$ & $0.55$\\
    \hline
   $T_\mathrm{c} $ & K & 28.4 & 18.2 & 12.5\\
   $\gamma_\mathrm{n}$ & mJ/mol\/K$^2$ & 28.0 & 20.2 & 19.2\\
   $\gamma_\mathrm{s} = \gamma _\mathrm{n} - \gamma _\mathrm{r}$ & mJ/mol\/K$^2$ & 23.3 & 14.0 & 13.8\\
   $\Delta C/T_\mathrm{c}$ & mJ/mol\/K$^2$ & 47.6 & 29.2 & 15.8\\
   $\mu_{0}{\left| dH_{c2}^{c}/dT \right|}$ & T/K & 2.13 & 0.96 & 0.59\\
   \footnote{from integration of $\Delta C$.}$\Delta F(0)$ & J/mol & 5.16 & 1.22 & 0.39\\
    \footnote{using $\Delta F(0)/V_\mathrm{m}=\mu_0 H_\mathrm{c}^2(0)/2$, with $V_\mathrm{m}$ from \cite{Rotter:2010jv}, using 122 formula unit for mol.}$\mu_{0} H_\mathrm{c}(0)$ & T & 0.468 & 0.230 & 0.130\\
    \footnote{assuming clean-limit WHH, $H_\mathrm{c2}^{c}(0)=0.73 T_{c}\left|{dH_{c2}^{c}/dT}\right|_{T_\mathrm{c}}$.}$\mu_{0}H_\mathrm{c2}^{c}(0)$ & T & 44.2 & 12.8 & 5.38\\
    \footnote{Density of states for both spin directions per 122 cell, using $N(\varepsilon_\mathrm{F}) = 3\gamma_\mathrm{n}/\pi^2 k_\mathrm{B}^2$.}$N(\varepsilon_\mathrm{F})$ & states/eV & 11.9 & 8.6 & 8.2\\
    \footnote{assuming $\Delta F = N(\varepsilon_\mathrm{F})\Delta^2(0)/4$.}$\Delta(0)$ & meV & 4.2 & 2.4 & 1.4\\
    \footnote{using $\mu_{0}H_{c2}^{c}(0)=\phi_{0}/{2\pi\xi_{ab}^2}(0)$.}$\xi_\mathrm{ab}(0)$ & nm & 2.73 & 5.08 & 7.82\\
    \footnote{using $\mu_0{\left|{dH_{c2}^{c}/dT}\right|}_{T_\mathrm{c}}=\phi_{0}/[{2\pi(0.74\xi_{0})^2}T_\mathrm{c}]$ from the clean-limit  GL relation $\xi_\mathrm{ab}^{-2}(T)|_{\mathrm{near}\,T_\mathrm{c}}=(0.74\xi_{0})^{-2}(1-T/T_\mathrm{c})$.}$\xi_\mathrm{0}$ & nm & 3.15 & 5.87 & 9.03\\
    \footnote{using $\mu_{0}H_{c}(0)=\phi_{0}/{2\sqrt{2}\pi\lambda_\mathrm{ab}(0)\xi_\mathrm{ab}(0)}$.}$\lambda_\mathrm{ab}(0)$ & nm & 182 & 200 & 229\\
    \footnote{using the clean-limit GL relation $\kappa_{c}(T_{c})=0.71\lambda_\mathrm{L}/0.74\xi_{0}$ with $\kappa_{c}(T_{c})$ obtained from the ratio of $dH_{\mathrm{c2}}^{c}/dT$ and $dH_{\mathrm{c}}/dT$, with $V_\mathrm{m}\mu_{0}\left|{dH_\mathrm{c}/dT}\right|_{T_c}^2=\Delta C/T_\mathrm{c}$ (Rutgers relation).}$\lambda_\mathrm{L}$ & nm & 156 & 165 & 212\\
    \footnote{assuming $\xi_0 = \hbar v_\mathrm{F}/\pi\Delta(0)$.}$v_\mathrm{F}$ & km/s & 63.8 & 68.0 & 60.7\\
    \footnote{assuming $\varepsilon_\mathrm{F}=k_\mathrm{B}T_\mathrm{F}=m^\star v_\mathrm{F}^{2}/2$}$T_\mathrm{F}$ & K & 602 & 492 & 373\\
  \end{tabular}
  \end{ruledtabular}
\end{table}

From specific heat, a wealth of information can be gained. The results are summarized in Table~\ref{Table1} together with assumptions and relations used. $T_\mathrm{c}$, $\Delta C/T_\mathrm{c}$, and the upper critical field slope $\mu_{0}{\left| dH_{c2}^{c}/dT \right|}$ are determined using an entropy-conserving construction, as shown in Fig.~\ref{Fig1}. By integrating $\Delta C(T)=C_\mathrm{e,s}-C_\mathrm{e,n}$, the condensation energy $\Delta F(0)$ is obtained. As seen in Table~\ref{Table1}, both $\Delta C/T_\mathrm{c}$ and $\Delta F(0)$ vary strongly with $T_\mathrm{c}$, with higher $T_\mathrm{c}$ samples also displaying higher $\Delta C/T_\mathrm{c}$ and $\Delta F(0)$. There is a clear increase of $\gamma_\mathrm{n}$ when approaching optimum doping, indicating an effective mass enhancement near $x_\mathrm{cr}$. From $\gamma_\mathrm{n}$ the density of states $N(\varepsilon _\mathrm{F})$ is found. An average zero-temperature superconducting gap  $\Delta (0)$ can then be estimated from $\Delta F(0)$. Both $N(\varepsilon _\mathrm{F})$ and $\Delta (0)$ increase with $T_\mathrm{c}$. The obtained $\Delta (0)$s are comparable to, but somewhat lower than those reported by angle-resolved photoemission spectroscopy (ARPES) for P-122 crystals with similar compositions \cite{Zhang:2012, Shimojima:2011, Yoshida:2014}. Attributing $\Delta F(0)$ only to a fraction $(\gamma_\mathrm{n}-\gamma_\mathrm{r})/\gamma_\mathrm{n}$ of the samples would increase the average $\Delta (0)$ by 10\%-20\%, largely eliminating the difference between our results and ARPES measurements. The ratios $\Delta (0)/k_\mathrm{B}T_\mathrm{c}$ are in line with measurements on Co-122 \cite{Hardy:2010hr}.

$\Delta F(0)$ leads us to the thermodynamic critical field $H_\mathrm{c}(0)$. The upper critical field $H_\mathrm{c2}(0)$ is estimated from its slope close to $T_\mathrm{c}$ and the Werthamer-Helfand-Hohenberg (WHH) relation \cite{Werthamer:1966}. This estimate gives good agreement with high-field transport measurements \cite{Analytis:2014}. Both critical fields reveal significant positive correlation with $T_\mathrm{c}$. Having the critical fields, the microscopic superconducting parameters, $\xi _\mathrm{ab}(0)$ and $\lambda _\mathrm{ab}(0)$ can be attained. They both decrease with increasing $T_\mathrm{c}$, but the decrease in $\lambda _\mathrm{ab}(0)$ is rather minor, as seen in Table~\ref{Table1}. Note that we make a distinction between $\lambda _\mathrm{ab}$ and $\lambda _\mathrm{L}$. $\lambda _\mathrm{L}$ is the zero-temperature London penetration depth while $\lambda _\mathrm{ab}=\lambda _\mathrm{ab}(T)$ is defined at all temperatures through $H_\mathrm{c}(T)$. Similarly, we obtain the constant $\xi_0$ from Ginzburg-Landau (GL) relations and relate $\xi_\mathrm{ab}(T)$ to $H_\mathrm{c2}(T)$ through $\mu_{0}H_\mathrm{c2}(T)=\phi_0/2\pi\xi_\mathrm{ab}^2(T)$. We do not make any explicit assumptions on the temperature dependence of $\kappa_\mathrm{c}(T) \equiv \lambda_\mathrm{ab}(T)/\xi_\mathrm{ab}(T)$.

\begin{figure}[b]
\begin{center}
\includegraphics[width=0.98\linewidth]{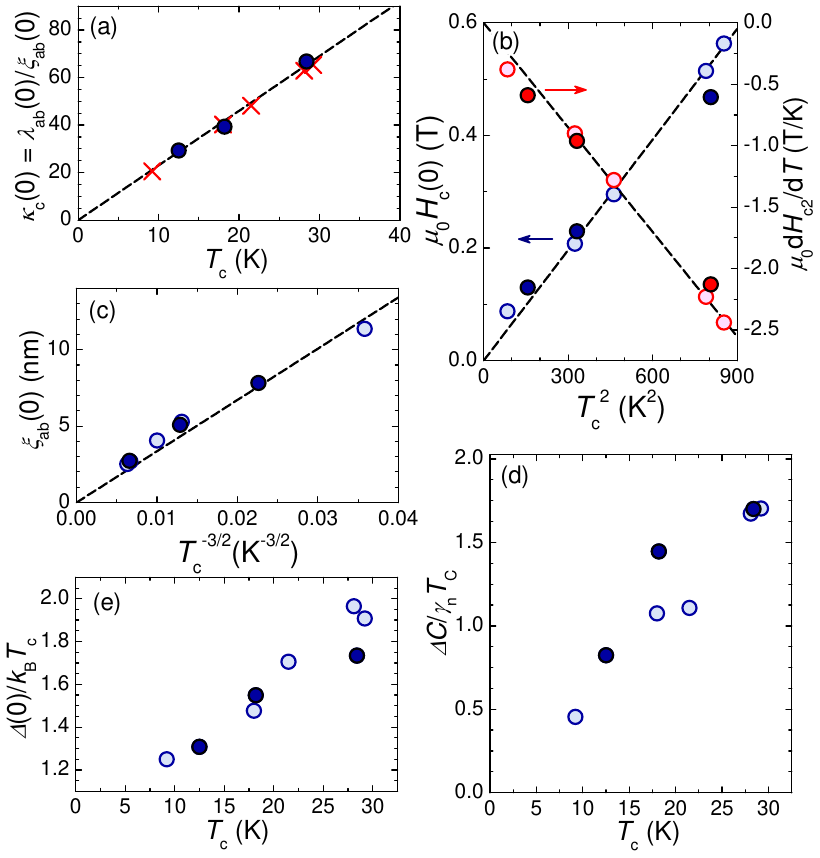}
\end{center}
\caption{(Color online) Approximate scaling dependence of superconducting parameters on $T_\mathrm{c}$. (a) $\kappa _\mathrm{c}$, (b) $H_\mathrm{c}$ and $\mathrm{d}H_\mathrm{c2}/\mathrm{d}T$,  (c) $\xi _\mathrm{ab}$, (d) $\Delta C/\gamma _\mathrm{n}T _\mathrm{c}$, (e) $\Delta /k _\mathrm{B}T_\mathrm{c}$. Filled symbols represent measurements of this work whereas partly filled symbols are from analysis of the data of Ref.~[\onlinecite{Chaparro:2012dw}]. For the latter, the additional assumption that $\kappa_\mathrm{c}\propto T_\mathrm{c}$ is made [crosses in (a)], as approximatively found for the present data. Dashed lines are linear fits through origin.}
\label{Fig2}
\end{figure}

In Fig.~\ref{Fig2}, we illustrate the variation of several superconducting parameters as a function of $T_\mathrm{c}$. The present measurements are also complemented with an analysis of the results of Ref.~[\onlinecite{Chaparro:2012dw}]. For this analysis, we estimate $H_\mathrm{c}(0)$ from the measured properties near $T_\mathrm{c}$ by assuming a scaling  $\kappa_\mathrm{c}(0) \propto T_\mathrm{c}$ that is roughly seen for the present data, as shown in Fig.~\ref{Fig2}(a). Figure~\ref{Fig2}(b) suggests that $H_\mathrm{c}(0)$ is largely proportional to $T_\mathrm{c}^{2}$, leading to an approximate scaling $\Delta F(0) \propto T_\mathrm{c}^{4}$. This observation is in drastic contrast to the expected $T_\mathrm{c}^{2}$ scaling of conventional superconductors, but follows the same trend as previously observed in a number of other Fe-based superconductors \cite{Xing:2014}. It is consistent with the BNC scaling, $\Delta C \propto T_\mathrm{c}^{3}$, provided that  $H_\mathrm{c}(T)$ has a normal BCS-like temperature dependence, in which case Rutgers relation would give $\Delta C/T \propto \left|{dH_\mathrm{c}/dT}\right|_{T_c}^2 \propto [H_\mathrm{c}(0)/T_\mathrm{c}]^2 \propto T_\mathrm{c}^2$.

The upper critical field slope $\mu _0dH_\mathrm{c2}^\mathrm{c}/dT$, shown on the right axis of Fig.~\ref{Fig2}(b), also reveals a $T_\mathrm{c}^{2}$ dependence, as reported in Ref.~[\onlinecite{Chaparro:2012dw}]. As a result, $\xi _\mathrm{ab}(0)$ varies approximately as $T _\mathrm{c}^{-3/2}$, as seen in Fig.~\ref{Fig2}(c). The ratio $\Delta C/\gamma _\mathrm{n}T_\mathrm{c}$, describing the relative specific heat discontinuity at $T_\mathrm{c}$, is shown in Fig.~\ref{Fig2}(d). Interestingly, this ratio is almost proportional to $T_\mathrm{c}$ rather than being constant, as it would be in the weak-coupling BCS theory ($\Delta C/\gamma _\mathrm{n}T_\mathrm{c}=1.43$). Thus, $\gamma_\mathrm{n}$ increases with $T_\mathrm{c}$ when approaching $x_\mathrm{cr}$, but not as fast as $\Delta C/T_\mathrm{c}$ does. The ratio of $\Delta(0)$ to $T_\mathrm{c}$ also shows an evident positive correlation with $T_\mathrm{c}$, see Fig.~\ref{Fig2}(e). However, this dependence is weaker, going from about 2 for the highest $T_\mathrm{c}$ and extrapolating to about 1 for $T_\mathrm{c} = 0$.

\begin{figure}[t!]
\begin{center}
\includegraphics[width=0.95\linewidth]{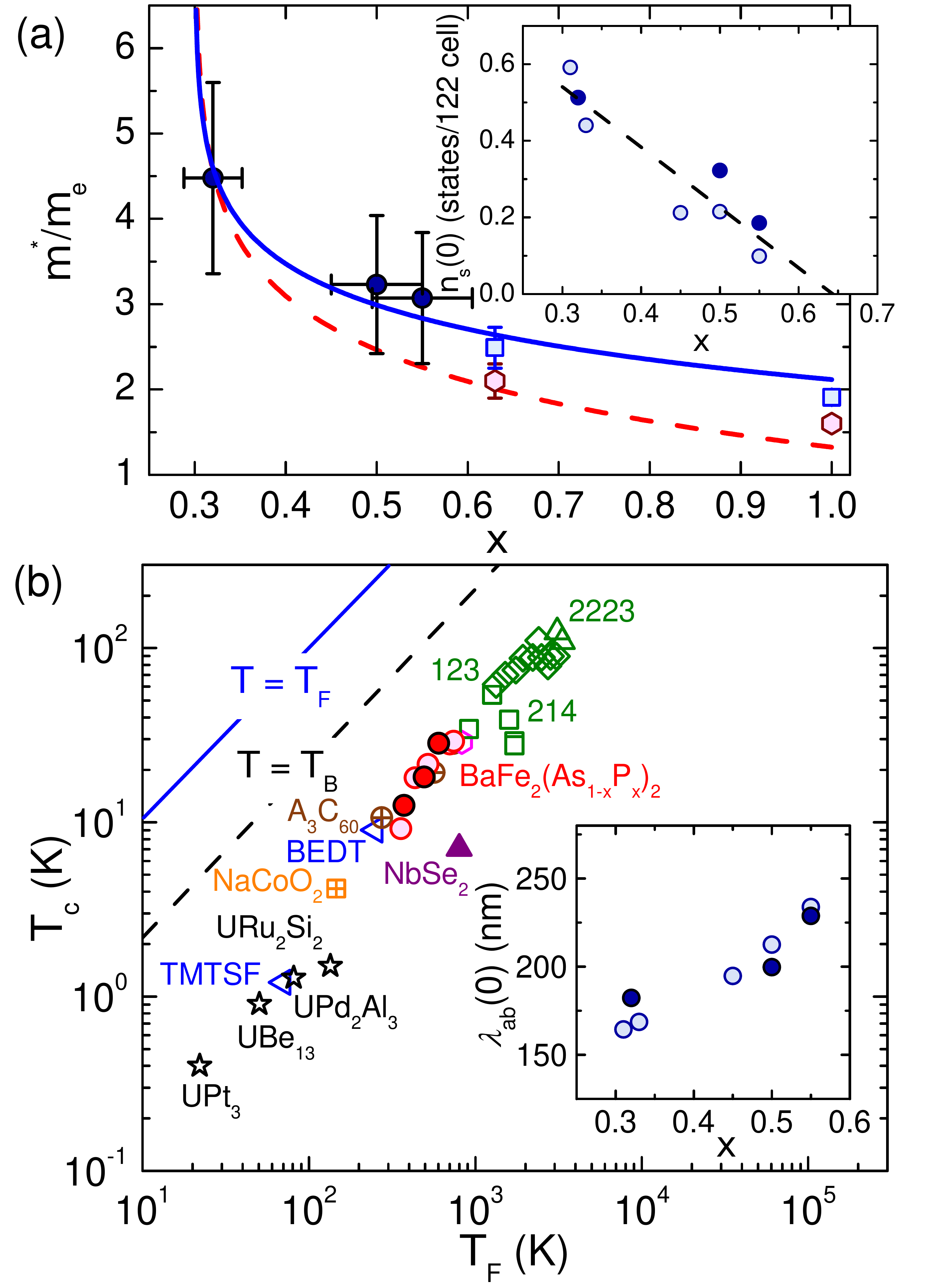}
\end{center}
\caption{(Color online) (a) Carrier effective mass as a function of $x$. Data points ($\varhexagon$) at $x = 0.63$ and $1.0$ are for the $\beta$ band from Refs.~[\onlinecite{Analytis:2010hl}] and [\onlinecite{Arnold:2011ip}]. They are shifted upwards ($\Square$) to account for the effective mass enhancement of all five Fermi sheets (see text). The dashed line is the best logarithmic fit to the effective mass enhancement measured by dHvA, $y = 1.0 -0.91\ln{(x - 0.30)}$ \cite{Walmsley:2013}. The solid line is a similar fit to our data and the two renormalized literature points, $y = 1.87 - 0.70\ln{(x - 0.30)}$. Inset: Number of superconducting carriers per 122 formula as a function of $x$. (b) Uemura plot showing the approximate linear scaling of $T_\mathrm{c}$ with $T_\mathrm{F}$. Data points other than those reported in this work were reproduced from Ref.~[\onlinecite{Hashimoto:2012}] and [\onlinecite{Uemura:2004}]. $T_\mathrm{B}$ is the Bose-Einstein condensation temperature for ideal 3D boson gases. Inset: Penetration depth $\lambda_\mathrm{ab}(0)$ as a function of $x$.}
\label{Fig3}
\end{figure}

From $\gamma_\mathrm{n}$, the carrier effective mass $m^\star$ is obtained. Band-structure calculations for fully doped BaFe$_{2}$P$_{2}$ give the bare band electronic specific heat $\gamma _\mathrm{b}= 6.94\,\mathrm{mJ\,mol}^{-1}\mathrm{K}^{-2}$ \cite{Walmsley:2013}. From $\gamma _\mathrm{n}/\gamma _\mathrm{b}$ we estimate an averaged effective mass $m^{\star}/m_\mathrm{e}$, taking into account the band mass $m_\mathrm{b}/m_\mathrm{e}$ of each of the five Fermi sheets and their respective contribution to the total density of states (DOS) \cite{Arnold:2011ip, Walmsley:2013}. The results are shown in Fig.~\ref{Fig3}(a) together with a logarithmic fitting function of the critical effective mass enhancement as previously applied to dHvA results \cite{Walmsley:2013}. Two well-documented data points are also included in the figure, one for the overdoped BaFe$_{2}$(As$_{0.37}$P$_{0.63}$)$_{2}$ \cite{Analytis:2010hl} and the other for BaFe$_{2}$P$_{2}$ \cite{Arnold:2011ip}, which have been confirmed by various separate studies \cite{Shishido:2010fn, Walmsley:2013}. The systematic vertical shift between our data and the previous results reflects different normalizations involved. From dHvA measurements \cite{Analytis:2010hl,Arnold:2011ip}, the $\gamma$ hole band is revealed to possess a significantly higher effective mass. Hence, $m^{\star}$ is shifted upwards when all bands are taken into account.

The relation $\mu_0 \lambda_\mathrm{L}^2=m^{\star}/n_\mathrm{s}e^2$ allows the evaluation of the superconducting carrier density $n_\mathrm{s}$ when $m^\star$ and $\lambda_\mathrm{L}$ are both known. In the inset of Fig.~\ref{Fig3}(a), $n_\mathrm{s}$ is plotted as a function of $x$. Evidently, $n_\mathrm{s}$ is displaying a possible vanishing around $x=0.65$, in the vicinity of which superconductivity is lost \cite{Nakajima:2013,Walmsley:2013}. On approaching $x_\mathrm{cr}$, $n_\mathrm{s}$ experiences a linear increase without critical signatures. We thus find that $n_\mathrm{s}$ increases with increasing $T_\mathrm{c}$ and condensation energy. Our data suggest a significant deviation from the assumption of $\Delta C/\gamma _{n}T_\mathrm{c} = 1.43$ used by Walmsley~\emph{et\,al.}~[\onlinecite{Walmsley:2013}]. We observed a clearly identifiable change in $\Delta C/\gamma _\mathrm{n}T_\mathrm{c}$, up to a factor of 3 - 4 as a function of $x$, see Fig.~\ref{Fig2}(d). This implies that an increase of $\Delta C/T_\mathrm{c}$ alone close to $x_\mathrm{cr}$ does not necessarily lead to a proportional increase of $m^\star$, as previously suggested in Ref.~[\onlinecite{Walmsley:2013}]. Such assumptions may instead lead to an overestimation of $\lambda_\mathrm{L}$ near $x_\mathrm{cr}$. In fact, we find a smooth, continuous decrease of $\lambda _\mathrm{ab}(0)$ with decreasing $x$, as shown in the inset of Fig.~\ref{Fig3}(b). This is in apparent contradiction with the results of Hashimoto \emph{et\,al.}~\cite{Hashimoto:2012}, where $\lambda_\mathrm{L}$ is reported to display a peak around $x_\mathrm{cr}$, increasing roughly by a factor 2 to 2.5. A possible explanation for this may be that specific heat mainly reflects carriers that are strongly gapped (hole band), while direct penetration depth measurements would weigh more heavily on carriers with long mean free paths but not necessarily all paired (electron band). This interpretation would be consistent with stronger inter-band coupling near $x_\mathrm{cr}$, where the system is more disordered, and weaker inter-band coupling but stronger gap anisotropy for the heavily overdoped, but cleaner system.

Besides $m^\star$, specific heat provides additional band-structure properties. We estimate the averaged Fermi velocity $v_\mathrm{F}$ from $\Delta (0)$ and $\xi _\mathrm{ab}(0)$, see Table~\ref{Table1}. The $v_\mathrm{F}$s obtained are of the same order of magnitude as those previously deduced through ARPES measurements \cite{Ye:2012} and change only little with $x$. This is largely supported by ARPES data, where a variation in $v_\mathrm{F}$ of less than 25\% was observed for a similar range of $x$. In Fig.~\ref{Fig3}(b) we visualize the variation of $T_\mathrm{c}$ as a function of $T_\mathrm{F}$ (obtained from $m^\star$ and $v_\mathrm{F}$) in the Uemura plot \cite{Uemura:2004, Hashimoto:2012}. Strikingly, our data fall seamlessly into the gap formed between cuprates and heavy Fermion compounds on the plot, showing an approximately linear scaling relation between $T _\mathrm{c}$ and $T_\mathrm{F}$. This strongly indicates that an electronic coupling mechanism may be involved in all these apparently very different superconducting materials. We note that this scaling relation works best as a function of $T_\mathrm{F}$. Plotted as a function of $\lambda_\mathrm{L}^{-2}$ or $n_\mathrm{s}$, P-122 would display a horizontal shift due to its multiple bands.

The Uemura plot here presents a rather different picture than that of Hashimoto \emph{et\,al.}~[\onlinecite{Hashimoto:2012}]. We find no anomaly in the superfluid density close to optimum doping and see a preserved scaling of $T_\mathrm{c}$ with $T_\mathrm{F}$. Parameters extracted from specific heat measurements predominantly represent portions of the Fermi surfaces with strong pairing. The non-vanishing residual electronic specific heat coefficient, also observed for instance in Co-122 \cite{Hardy:2010hr} indicate that patches of the Fermi surface are only very weakly paired or remain unpaired. While regular disorder could possibly cause this, it seems unlikely for the case of phosphorous-doped 122, where at least the strongly overdoped side is in the clean limit. A more likely explanation is a decreasing interband coupling with increasing doping $x$, causing both increased gap anisotropy and weakened superconductivity.

In conclusion, we find that a higher Fermi temperature is beneficial rather than detrimental for the electronically mediated superconductivity in phosphorous-doped 122. With increasing overdoping, superconductivity is suppressed through a decreasing superconducting carrier density.

\begin{acknowledgments}
We are grateful for equipment supported by the K.\,\&\,A.\ Wallenberg foundation.
Work at Argonne was supported by the U.S. Department of Energy, Office of Science, Basic Energy Sciences, Materials Sciences and Engineering Division.
\end{acknowledgments}

\end{document}